\documentclass{emulateapj}
\usepackage{color}

\shorttitle{Suzaku Observation of 4U 1630$-$47}
\shortauthors{Hori et al.}
\begin{document}

\title{Suzaku Observation of the Black Hole Binary \\ 4U 1630--47 in the Very High State}

\author{TAKAFUMI HORI\altaffilmark{1}, YOSHIHIRO UEDA\altaffilmark{1}, MEGUMI SHIDATSU\altaffilmark{1}, TAIKI KAWAMURO\altaffilmark{1}, AYA KUBOTA\altaffilmark{2}, CHRIS DONE\altaffilmark{3}, SATOSHI NAKAHIRA\altaffilmark{4}, KOHJI TSUMURA\altaffilmark{5}, MAI SHIRAHATA\altaffilmark{6,7}, and TAKAHIRO NAGAYAMA\altaffilmark{8}}

\email{hori@kusastro.kyoto-u.ac.jp}

\altaffiltext{1}{Department of Astronomy, Kyoto University, Kitashirakawa-Oiwake-cho, Sakyo-ku, Kyoto 606-8502, Japan}
\altaffiltext{2}{Department of Electronic Information Systems, Shibaura Institute of Technology, 307 Fukasaku, Minuma-ku, Saitama 337-8570, Japan}
\altaffiltext{3}{Department of Physics, University of Durham, South Road, Durham, DH1 3LE, UK}
\altaffiltext{4}{ISS Science Project Office, Institute of Space and Astronautical Science (ISAS), Japan Aerospace Exploration Agency (JAXA), 2-1-1 Sengen, Tsukuba, Ibaraki 305-8505, Japan}
\altaffiltext{5}{Frontier Research Institute for Interdisciplinary Science, Tohoku University, Sendai, Miyagi, 980-8578, Japan}
\altaffiltext{6}{National Institutes of Natural Science
National Astronomical Observatory of Japan (NAOJ)
2-21-1 Osawa, Mitaka, Tokyo 181-8588, JAPAN}
\altaffiltext{7}{Department of Space Astronomy and Astrophysics, Institute of Space and Astronautical Science (ISAS) Japan Aerospace Exploration Agency (JAXA) Sagamihara, Kanagawa 252-5210, Japan}
\altaffiltext{8}{Department of Astrophysics, Nagoya University, Chikusa, Nagoya 464-8602, Japan}

\begin{abstract}
We report the results from an X-ray 
and near-infrared
observation of the Galactic black hole binary 4U 1630--47
in the very high state, performed with {\it Suzaku} 
and IRSF 
around the peak of the 2012 September-October outburst. 
The X-ray spectrum is approximated by a steep power law, with photon
index of 3.2, identifying the source as being in the very high state. A
more detailed fit shows that the X-ray continuum is well described by
a multi-color disc, together with thermal and non-thermal
Comptonization. The inner disc appears slightly truncated by
comparison with a previous high/soft state of this source, even taking
into account energetic coupling between the disc and corona, although
there are uncertainties due to the dust scattering correction.
The near-infrared fluxes are higher than the extrapolated disc model,
showing that there is a contribution from irradiation in the outer disk and/or the
companion star at these wavelengths.
Our X-ray spectra do not show the Doppler shifted iron emission lines
indicating a baryonic jet which were seen four days previously in an
XMM-Newton observation, despite the source being in a similar
state. There are also no significant absorption lines from highly
ionized irons as are seen in the previous high/soft state data. We show
that the increased source luminosity is not enough on its own to make
the wind so highly ionized as to be undetectable. This shows that the
disc wind has changed in terms of its launch radius and/or density
compared to the high/soft state.
\end{abstract}

\keywords{accretion, accretion disks --- black hole physics --- stars: individual (4U 1630--47) --- X-rays: binaries}

\section{Introduction}\label{intro}

Galactic Black hole binaries (BHBs) are ideal objects to study the
physics of accretion flow onto a black hole. 
BHBs take several different states with distinct spectral
and timing properties \citep[for a review see, e.g.,][]{Done07}. 
Generally, when the mass accretion rate (and hence the luminosity)
is low, a BHB exhibits the Low/Hard State (LHS), where the spectrum is
dominated a power law with a photon index of $\Gamma \sim 1.7$ and a
high energy cut-off at $\sim 100$ keV. At a higher mass accretion rate,
the spectrum is dominated by optically thick, thermal emission from the
accretion disk, which is well represented by the Multi-Color Disk (MCD)
model \citep{Mitsuda84,Makishima86}. This state is called the High/Soft
State (HSS). When the mass accretion rate becomes even larger and the
luminosity is close to the Eddington limit, a BHB takes the so-called
Very High State (VHS), where the spectra are characterized by strong
Comptonization. Although the VHS is a key state to understand the
accretion disk physics under very high mass accretion rates, it has been
rather poorly investigated because it is quite rare that a BHB takes
this state.

The evolution of the state of a BHB during an outburst is 
complex, and not solely determined by
mass accretion rate. In typical outbursts, a
BHB shows hysteresis with a q-shaped curve on a hardness-intensity diagram 
\citep[e.g.,][]{Fender04}. Sources start from the LHS,
then make a transition into the HSS, sometimes via the VHS.
The source remains in the HSS while the luminosity declines, and then turns back to the
LHS, at a luminosity which can be a factor of 2-20 times smaller than the initial 
LHS-HSS transition.
This hysteresis behavior probably originates from the accretion rate changing faster
than the viscous timescale in the geometrically thin disk \citep{Gladstone07}.
However, BHBs in outbursts do
not always exhibit similar evolutional tracks on this diagram 
(e.g., \citealt{Nakahira14}),
implying that the physical processes governing the
spectral evolution in BHBs may be even more complicated. 

An important issue in understanding black-hole accretion flow in
each state is whether the standard disk extends down to the
innermost stable circular orbit (ISCO) around the black hole 
or is truncated at a larger radius.
In the MCD model, the disk bolometric luminosity $L_{\rm disk}$
is related to the disk temperature $T_{\rm in}$ and innermost disk
radius $r_{\rm in}$ as $L_{\rm disk} = 4 \pi r_{\rm in}^2 \sigma T_{\rm
in}^4$, where $\sigma$ is Stefan-Boltzmann constant.
The standard disk is believed to extend to the ISCO in the HSS, since
the observed $r_{\rm in}$ value of a BHB is found to be constant over a
wide range of $L_{\rm disk}$ \citep[e.g.,][]{Ebisawa93}.
By contrast, many studies suggest that the accretion disk in the LHS is
truncated before reaching the ISCO \citep[e.g.,][]{Makishima08,
Tomsick09,Shidatsu11,Shidatsu13} and the inner part is probably 
replaced by Radiative Inefficient Accretion Flow (RIAF) 
\citep[e.g.,][]{Esin97}.
The geometry of the innermost part of the accretion flow 
in the VHS is less clear, however. 
This is because in this state a hot corona might be 
strongly Compton-scattering photons emitted from the standard disk
\citep{Zdziarski02,Gierlinski03,Kubota04b}, making it 
difficult to directly measure the intrinsic disk parameters.
There are some reports that the innermost disk radius in the VHS is
somewhat larger than that in the HSS
\citep{Kubota04b,Done06,Tamura12}, indicating that the disk
may be truncated before the ISCO. It is crucial to establish this
picture by using high-quality broad band X-ray spectra of BHBs in the
VHS.

4U 1630--47 is identified as a Galactic BHB because of the similarity of
the spectral and temporal behaviors to other BHBs, although the mass of
the compact object has not been measured yet. This object exhibits
quasi-periodic outbursts with 600--700 day cycles. This period is
very rapid as a BHB, probably indicating a high time-averaged mass
accretion rate. The X-ray spectra of 4U 1630--47 are heavily absorbed
\citep[$N_{\rm H} \sim 8\times 10^{22}$ cm$^{-2}$;][]{Kubota07}, suggesting 
that the object is located near the Galactic center, although the 
accurate source distance $D$ is unknown. The nature of the companion 
has not been firmly identified yet; \citet{Augusteijn01} infer that 
it may be an early type star by assuming that the whole $K$-band magnitude 
in outburst comes from the companion.

Outbursts of 4U 1630--47 are observed with many missions since 1980s,
including {\it EXOSAT} \citep{Parmar86}, {\it Einstein/Ginga/ASCA} 
\citep{Parmar97}, {\it Beppo-SAX} \citep{Oosterbroek98}, 
{\it RXTE} \citep[e.g.,][]{Tomsick98,Kuulkers98,Tomsick00,Trudolyubov01,Abe05},
{\it INTEGRAL} \citep{Tomsick05}, {\it Suzaku} \citep{Kubota07}, 
{\it XMM-Newton} \citep{Diaz Trigo13},
and {\it NuSTAR} \citep{King14}. 
By fitting the RXTE spectra taken between 2002 and 2005 with the MCD
plus power-law model, \citet{Tomsick05} suggest that the accretion disk
exhibited a state transition between the standard disk and the slim disk
at high mass accretion rates. Similar results are also reported by
\citet{Abe05} using the RXTE data. Most of these studies, however, apply
simple, phenomenological models to the spectra that are limited in the
energy resolution or band coverage. To understand the physical
conditions of the accretion flow in the VHS, detailed analysis based on
physically self-consistent models would be required.

4U 1630--47 is an important target to study the disk wind and jet
formation mechanism. \citet{Kubota07} discover absorption lines from
highly ionized iron ions in the {\it Suzaku} spectra of 4U 1630--47 in
the HSS, indicating the presence of disk wind, at least, in this
state. Recently, \cite{King14} also report an absorption-line feature
around 7 keV when the source was in a very similar state to that in the
first observation of \citet{Kubota07}. These facts suggest a relatively
high inclination angle of the system, $i \sim 70^\circ$, from analogy
with other BHBs with disk wind signatures
\citep{Ueda98,Kotani00,Ponti12}.
This is also supported by \citet{Munoz-Darias13} from the rather hard
color of the disc in the HSS, which can be explained by relativistic effects on the disc 
spectrum at high inclination.
\citet{Diaz Trigo13} discover Doppler shifted emission
lines of relativistic baryonic jets from an {\it XMM-Newton} observation
performed on 2012 September 28.

In this paper, we analyze the {\it Suzaku} data of 4U 1630--47 taken in
the VHS during the 2012 outburst. The observations and data reduction
are described in Section \ref{obs}. We present the analysis results of
the VHS spectra in Section \ref{analysis}. The method to correct for the
dust scattering effects is explained there. The results of re-analysis
of the HSS spectra taken in 2006 February are presented in Section
\ref{analysis2}. We report the results from infrared observations
performed one day before the {\it Suzaku} one at the Infra-Red Survey
Facility (IRSF) in Section~\ref{infrared}. Discussions are given in
Section \ref{discuss}, focusing on the inner disk truncation, jet
lifetime, and disk wind. Throughout the paper, we adopt the distance of
$D=10$ kpc and inclination angle of $i=70^{\circ}$ as the reference
values. We assume the Solar element abundances given by
\citet{Anders89}. All errors attached to spectral parameters are those
at 90\% confidence limits for a single parameter of interest.

\section{Observation and Data Reduction}\label{obs}

\begin{figure*}[t]
\epsscale{1.0}
\plotone{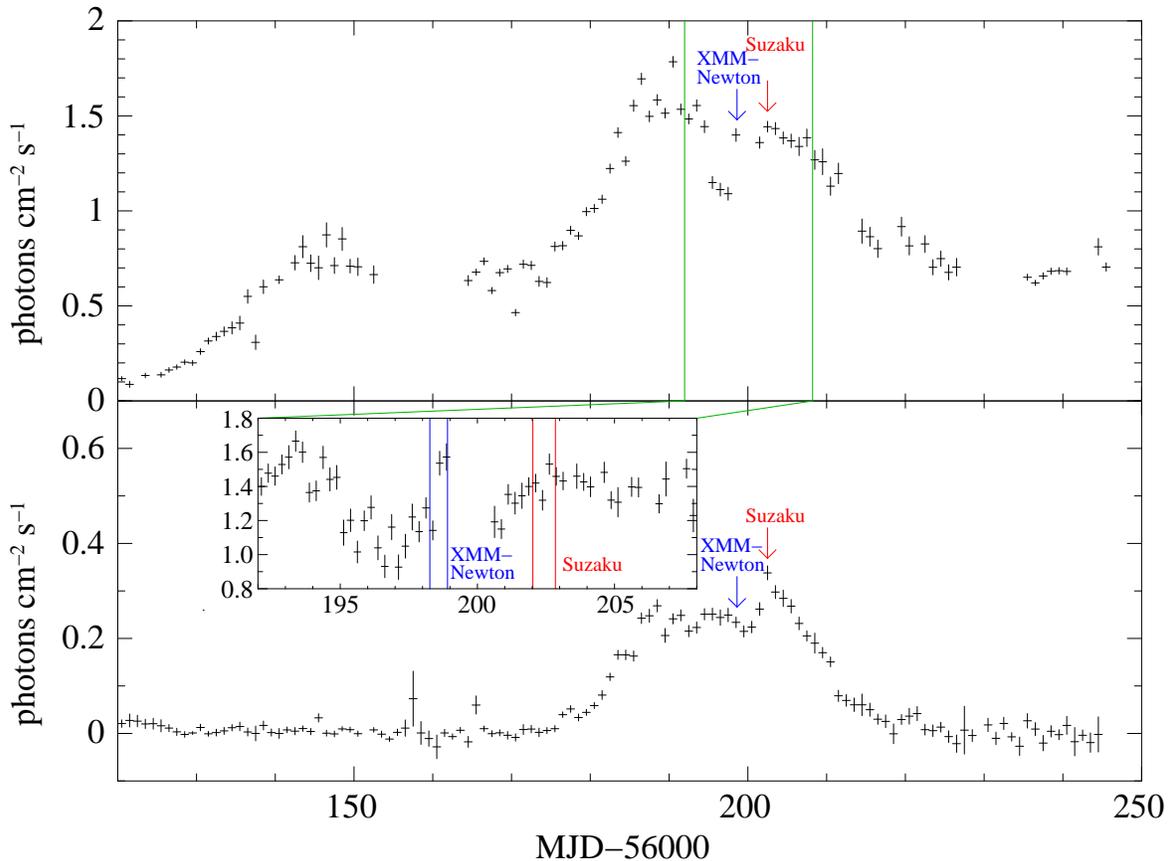}
\caption{Long term light curves of 4U 1630--47 in the 3--10 keV and 
15--50 keV bands obtained with {\it MAXI}/GSC (top panel) and {\it Swift}/BAT 
(bottom panel), respectively. 
The photon fluxes are converted from the count rates by assuming a
power-law photon index of 2.1.
The attached error bars correspond to the statistical errors
($1\sigma$) only. Systematic errors of $\sim$10\% could be present in
the {\it MAXI}/GSC light curve.
The observation epoch by {\it Suzaku} (this work) and that 
by {\it XMM-Newton} \citep{Diaz Trigo13} are indicated by the red 
(right) and blue (left) down-arrows, respectively.
The inset plots the {\it MAXI} light curve in 1/4 day bins
around the {\it Suzaku} observation. The red and blue lines represent the observation
epoch by {\it Suzaku} and {\it XMM-Newton}, respectively. \\
\label{MAXI_BAT}}
\end{figure*}

\begin{figure*}
\begin{center}
\epsscale{1.0}
\plottwo{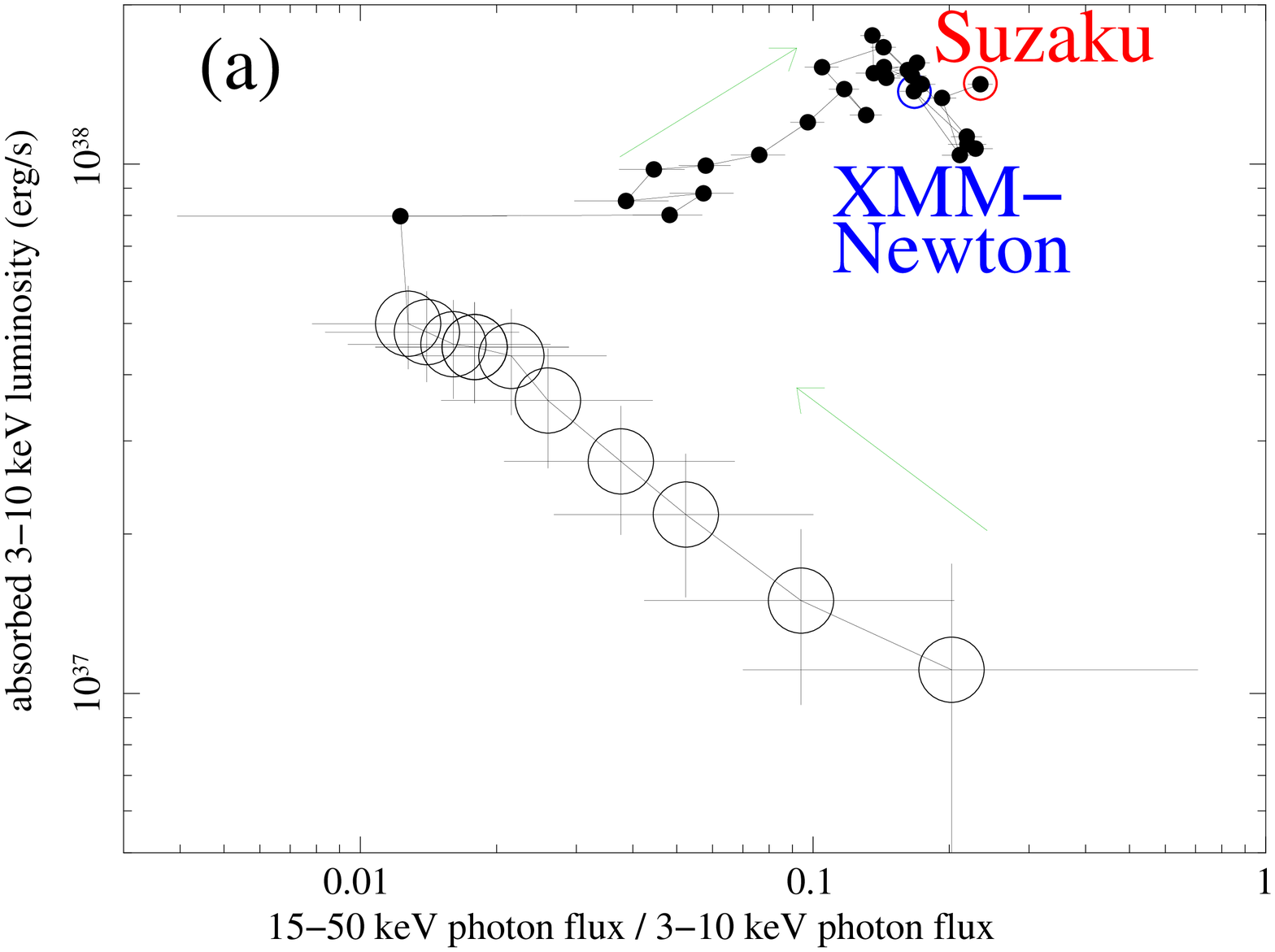}{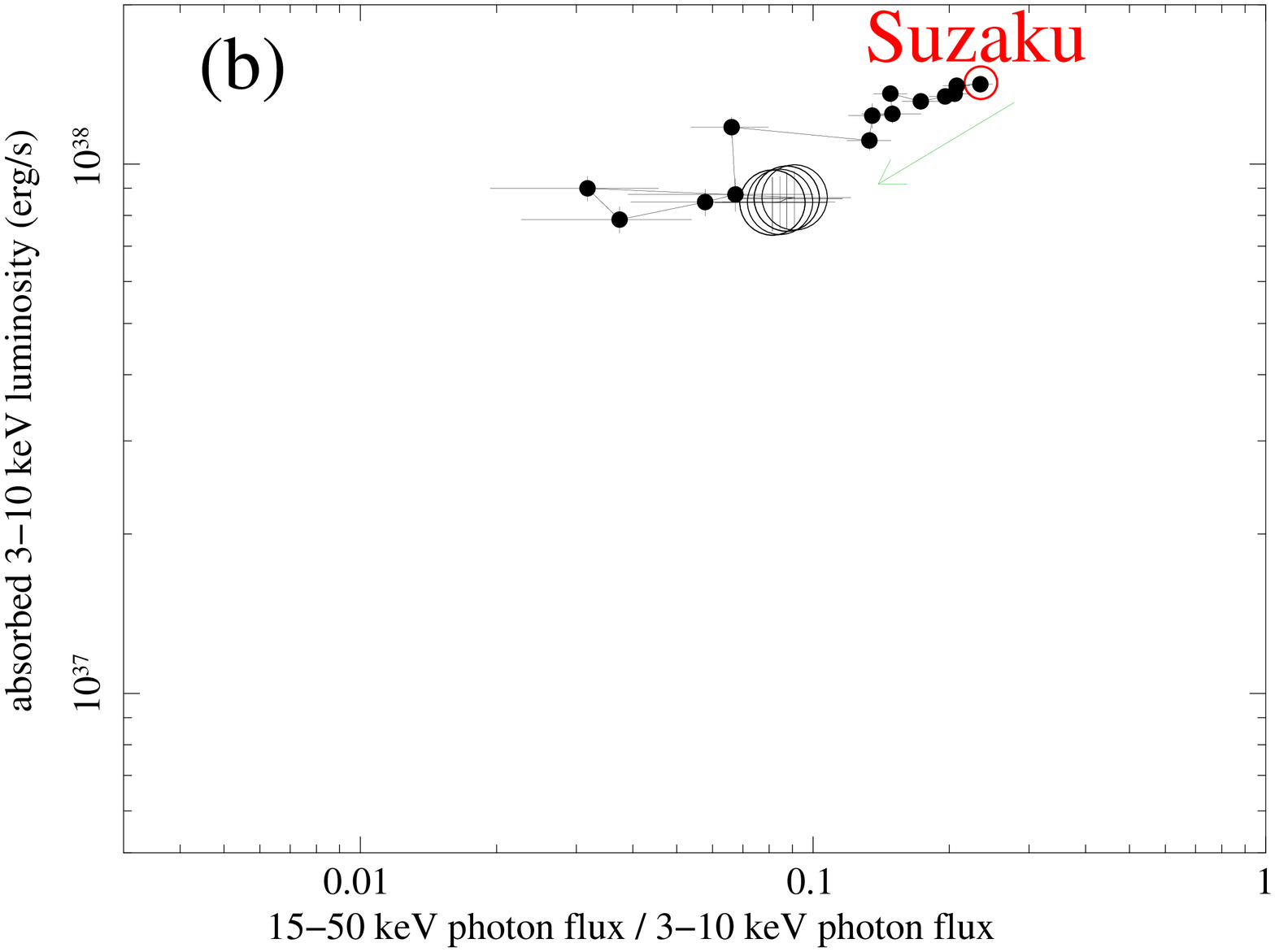}
\caption{The hardness-intensity diagrams using the {\it MAXI} and {\it Swift}/BAT
light curves shown in Figure~\ref{MAXI_BAT} (MJD=56120-56250). Panels (a) and
(b) correspond to that before and after the {\it Suzaku} observation, 
respectively. The epoch of the {\it XMM-Newton} ({\it Suzaku})
observation is marked by the blue (red) circle. To reduce the
statistical errors, the light curves are binned in 5 days during
MJD=56120$-$56175 and MJD=56220$-$56250, whose data points are indicated by the open
circles. The other data (MJD=56176$-$56219) are binned in 1 day, 
indicated by the small filled circles.
Starting from the bottom part in panel (a), it evolves in clockwise
direction, as shown by the green arrows. After the {\it Suzaku}
observation, it moves to the lower left, as shown by the green arrow in
panel (b). \label{hardness}}
\end{center}
\end{figure*}

We performed a ToO (Target of Opportunity) observation of 4U 1630--47
with {\it Suzaku} from 2012 October 2 00:24:59 (UT) to 20:38:24. Figure
\ref{MAXI_BAT} shows its long-term light curves in the 3--10 keV and
15--50 keV bands obtained with the Gas Slit Camera (GSC) on-board
Monitor of All-sky X-ray Image \citep[{\it MAXI};][]{Matsuoka09} and
the Burst Alert Telescope (BAT) on-board {\it Swift} \citep{Burrows05},
respectively. 
The {\it MAXI} light curves are obtained by the image
fitting method described in \citet{Hiroi13}.
The epoch of the {\it Suzaku} observation is indicated
with the down arrow (left) in Figure~\ref{MAXI_BAT}. As noticed from the
figure, it corresponds to the peak of the hard X-ray flux above 15 keV
during the 2012 September-October outburst. Detailed behavior in the
soft band ($<10$ keV) was more complex, showing large variability for
$\sim$10 days before the Suzaku observation.
The inset in Figure~\ref{MAXI_BAT}
plots the {\it MAXI} light curve in 1/4 day bins around the 
{\it Suzaku} observation. The data between MJD = 56199--56200.25 are not
available because the source was out of the field-of-view of {\it MAXI}.
Figure~\ref{hardness} plots the hardness-intensity
diagram corresponding to the epoch shown in Figure~\ref{MAXI_BAT}. 
It first evolved in the opposite direction to the
normal track of BHBs in outbursts \citep{Fender04},
i.e. it moved from left to right along the upper branch of the hardness-intensity diagram.
We note that this is 
a well known behavior of the VHS in several BHBs \citep[e.g.,][]{Tomsick05,Motta12}.

{\it Suzaku} \citep{Mitsuda07} carries two types of instruments, the X-ray
Imaging Spectrometer (XIS) and the Hard X-ray Detector (HXD). The XIS is
a focal plane X-ray CCD camera coupled to the X-ray telescope (XRT) and
covers the 0.2--12 keV band. At the time of our observation, two
front-side illuminated XISs (FI-XISs), XIS0 and XIS3, and one back-side
illuminated XIS (BI-XIS), XIS1, were available. The HXD is a collimated
type non-imaging instrument, consisting of 
silicon PIN diodes and gadolinium
silicon oxide (GSO) crystal scintillators, which cover the 10--70 keV and 40--600 keV
band, respectively. To minimize photon pile-up events, all XISs were
operated with a burst option of 0.1 sec exposure per frame. The 1/4
window mode was employed for XIS0 and XIS1, while the full window mode
is used for XIS3. The net exposure time for XIS0 and XIS1 is $\approx$3
ks and that for XIS3 is $\approx$0.5 ks. The HXD was operated in the
standard mode, and its net exposure time corrected for dead time is
$\sim$40 ks.

We utilize the ``cleaned'' event files reduced by the pipeline
processing version 2.8.16.34, and analyze them in a standard manner with
HEASOFT version 6.12 and Calibration Database (CALDB) released on 2012
September 2. To extract the XIS light curves and spectra, we accumulate
photon events in a circular region with a radius of $1^{\prime}.8$
centered at the source position. The background of the XIS is neglected
because the contribution is less than 1\% of the source counts. As for
the background of the HXD, we utilize the so-called ``tuned'' background
files provided by the {\it Suzaku} team. The modeled cosmic X-ray background
(CXB) is further subtracted from the PIN data but not from the GSO data
since its contribution is less than 0.1$\%$ of the total GSO background.

\begin{figure}[h]
\epsscale{1.0}
\plotone{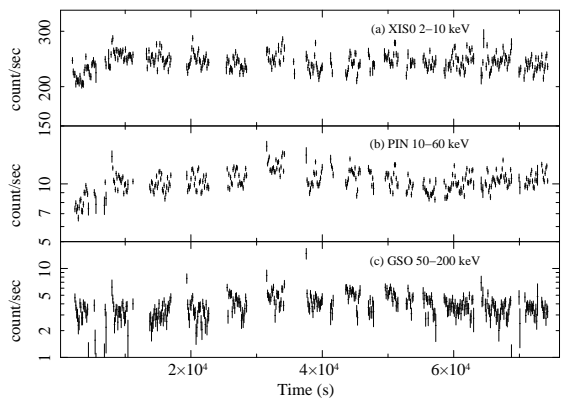}
\caption{{\it Suzaku} light curves of 4U 1630--47 binned in 128s obtained with 
XIS0 in the 2--10 keV band (top panel), HXD/PIN in the 10--60 keV band 
(middle panel), and HXD/GSO in the 50--200 keV band (bottom panel). The
background subtraction and the dead time correction are applied for the
HXD data.\label{4U_lc}}
\end{figure}

\begin{figure}[h]
\epsscale{1.0}
\plotone{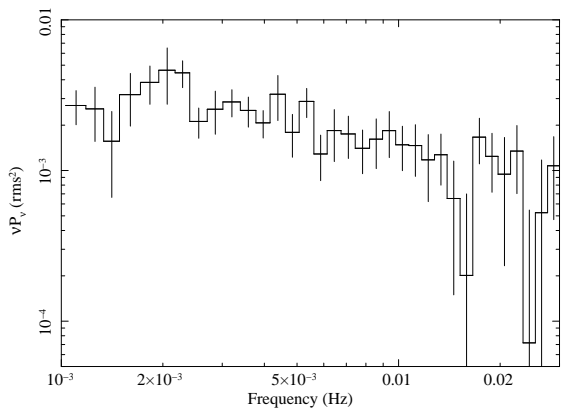}
\caption{Normalized power spectral density calculated from the HXD/PIN 
light curve in the 10--60 keV band. The white noise is subtracted.\label{PSD}}
\end{figure}

Figure~\ref{4U_lc} plots the {\it Suzaku} light curves of 4U 1630--47 in three
energy bands, 2--10 keV (XIS), 10--60 keV (PIN), and 50--200 keV
(GSO). The averaged 2--10 keV flux measured by XIS0 is about $1.1 \times 10^{-8}~{\rm erg~s^{-1}~cm^{-2}}$. The PIN and GSO light curves are corrected for dead time. 
Figure~\ref{PSD} shows the normalized power spectral density (PSD)
obtained from the PIN light curve with 0.001 sec binning. This
PSD is calculated over the entire observation duration by
using the powspec tool in the XRONOS software package
version 5.22.
As noticed, no significant quasi
periodic oscillation (QPO) is detected in the range of 0.001--0.02 Hz.

The time-averaged spectra of the XIS and the HXD are plotted in the
uppermost panel of Figure~\ref{4U_spec}. We create the response matrix
files and auxiliary response files of the XIS with the FTOOL {\tt
xisrmfgen} and {\tt xissimarfgen} \citep{Ishisaki07}, respectively. 
As for the response file of the HXD, we use {\tt
ae\_hxd\_pinxinome11\_20110601.rsp} for PIN, and {\tt
ae\_hxd\_gsoxinom\_20100524.rsp} and {\tt
ae\_hxd\_gsoxinom\_crab\_20100526.arf} for GSO. To account for possible
calibration uncertainties, 1$\%$ systematic errors are included in each
spectral bin of all spectra. 
Considering possible calibration uncertainties in 
the low energy tail in the pulse height amplitude distribution,
we increase the errors to 3\% at energies below 
1.7 keV in the XIS spectra.

In order to compare our results in the VHS with those in the HSS, we
re-analyze the XIS spectra taken between 2006 February 8 15:09:56 to
February 9 03:20:18 (ObsID=400010010) published by \citet{Kubota07}. 
The main reason is to obtain the MCD parameters in the HSS by taking 
into account dust-scattering effects, which were neglected in 
\citet{Kubota07}. All XISs were operated with the 1/4 window mode 
and 1-s burst option in this observation. The net exposure is 
22 ks for each XIS. The XIS spectra are 
integrated within a circular region of a radius of $4^{\prime}.3$, while 
central region of $30^{\prime \prime}$ are excluded in order to minimize the 
effect of pileup. The HXD was operated in the
standard mode, and its net exposure time corrected for dead time is
$\sim$22 ks. The same systematic errors as those applied for the 2012
data are included. The flux in
the 2--10 keV band is about $6.6 \times 10^{-9}~{\rm erg \ s^{-1} \
cm^{-2}}$.

\section{Analysis of the Time-Averaged Spectra in the Very high state}\label{analysis}

In this section, we present the results of detailed spectral analysis
of the {\it Suzaku} data of 2012 October. The spectral fit is
performed by using XSPEC version 12.8.0. We simultaneously fit the
spectra of three XISs, HXD/PIN, and HXD/GSO, using the energy range of
1.2--9.0 keV for XIS0 and XIS3, 1.2--8.0 keV for XIS1, 13--60 keV for
PIN, and 7 0--200 keV for GSO. The XIS data in the 1.7--1.9 keV energy
band were excluded in the spectral fit to avoid the large calibration
uncertainties. The cross normalization factor between XIS0 and HXD
(both PIN and GSO) is fixed at
1.164\footnote{http://www.astro.isas.ac.jp/suzaku/doc/suzakumemo/\\suzakumemo-2008-06.pdf}
while those between XIS0, XIS1, and XIS3 are set to be free. We extend
the energy range of the spectral fit to 0.01--1000 keV when using
convolution models.

\subsection{Effects by Dust Scattering}\label{dust}

Because 4U 1630--47 suffers from heavy absorption ($N_{\rm H} \sim
8\times10^{22}$ cm$^{-2}$) by the interstellar medium, we have to take
into account the effects by dust scattering, which become more
significant at lower energies. A part of the direct emission from the
object is scattered out by interstellar dust, while photons emitted
toward slightly different angles are scattered in the line of sight,
thus making the dust-scattering halo around the point source. When one
cannot integrate all the emission including both direct and halo
components, the scattered-in and scattered-out photons are not canceled
each other. The situation becomes more complex for a variable source
because there is a time delay between the scattered-in and direct
photons due to their different path lengths. Hence, we have to take into
account these effects in the spectral analysis of a heavily absorbed
object unless the spectrum is constant {\it and} the integrated image
region is sufficiently large to fully cover the dust-scattering halo.

Following the work by \citet{Ueda10}, who applied the same procedure
to {\it Suzaku} data of GRS 1915+105, we first estimate the fraction
of dust scattering halo of 4U 1630-47 contained in our XIS
spectra. According to \citet{Smith02}, who studied the detailed image
profile of the dust scattering halo of the bright Galactic source GX
13+1 observed with Chandra, the surface density of the halo profile
can be approximately represented by a power law with $r^{-1.7}$ below
$r=2^{\prime}.5$ and $r^{-2.8}$ above $r=2^{\prime}.5$. We then run
the FTOOL {\tt xissimarfgen} by assuming this profile and find that
about 65\% of the total photons of the dust-scattering halo is
included in the VHS spectra within our integration region. In spectral
fit, we utilize the {\tt Dscat} model, a local model on XSPEC
developed by \citet{Ueda10}. It calculates the fractions of the
scattered-in and scattered-out components as a function of energy, by
referring to the cross section of the dust scattering given by
\citet{Draine03}\footnote{We adopt that for the Milky way with $R_{\rm
V}=3.1$.}. The free parameters of the {\tt Dscat} model are the
hydrogen column density ($N_{\rm H}$) and the fraction of the halo
(scattered-in) component contained in the XIS spectra (hereafter
``scattering fraction''), which is fixed at 0.65. This is not applied
to the spectra of the HXD, which has a sufficiently large field of
view. For simplicity, we neglect the effect of time variability. 
The maximum time delay of the scattered-in component to the
direct one is estimated to be $\approx$1.5 days at the outer
integration radius of $1^{\prime}.8$ by assuming that the scatterers
are located at the half distance of $D=10$ kpc. 
The {\it MAXI} light curve (Figure \ref{MAXI_BAT})
shows that the soft X-ray flux was constant within 10\% for 1.5 days
before the {\it Suzaku} observation. We confirm that our conclusions are
not affected (see Section~\ref{truncation}).

\begin{figure*}[t]
\epsscale{1.0}
\plotone{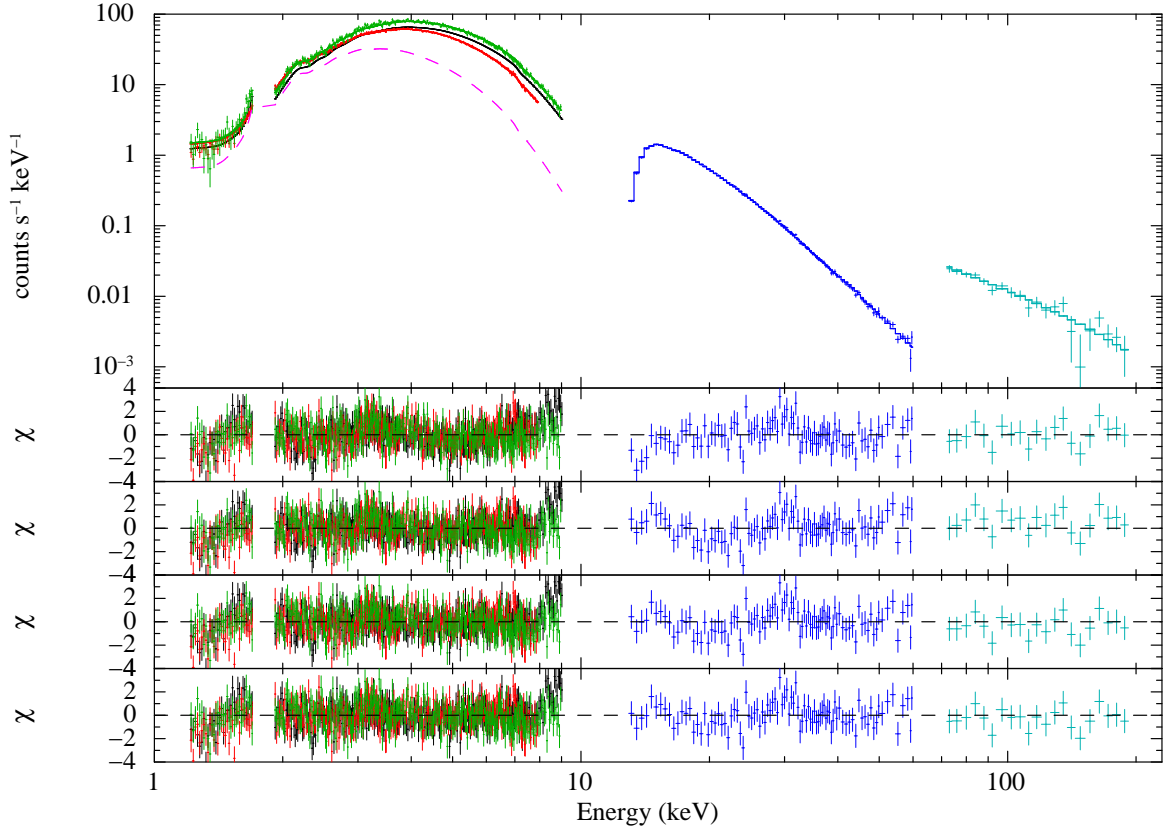}
\caption{Time averaged spectra of the XIS0 (black), XIS1 (red), 
XIS3 (green), PIN (blue), and GSO (sky blue) spectra (top panel). 
The magenta dashed curve indicates the estimated contribution from 
the dust scattering halo in XIS0 (see text).
(a)--(d) Residuals between the data and model in units of $\chi$ for
Models~A, B, C, and D, from upper to lower.\label{4U_spec}}
\end{figure*}

In the spectral fit presented below, we multiply to an intrinsic model
both the {\tt wabs} and {\tt Dscat} models, which take into account
the interstellar photo-electric absorption \citep{Morrison83} and
dust-scattering effects, respectively. 
The hydrogen column density of the {\tt Dscat} model
is linked to that of the {\tt wabs} model.
The estimated
contribution of the halo component in the XIS0 spectrum is plotted by
the magenta curve in the upper limit of Figure~\ref{4U_spec}. We find
that the net effects by dust scattering are $\approx$83\% at 2 keV and
$\approx$66\% at 3 keV compared with the case when neglected. Thus,
consideration of the dust-scattering effects is critical when we
discuss the intrinsic source spectrum at low energies.

\subsection{Simple Models (Model~A)}

To begin with, we try simple phenomenological models to fit the XIS+HXD
spectra in the VHS. We find a single power law model gives a very poor
fit ($\chi^2/{\rm dof}=4386/1032$) with the best-fit photon index of
$\Gamma \simeq 3.2$. The model leaves systematic hard X-ray excess above
$\approx$20 keV that monotonically increases with energy and soft excess
below $\approx$2 keV. The residuals indicate that the overall spectral
shape is approximated by a power law flatter than $\Gamma \simeq 3.2$ on which
an additional soft component is required.

Next, we fit the spectra with a MCD plus a power law model ({\tt
diskbb + pow} in the XSPEC terminology, hereafter referred as
``Model~A''). This is a conventional model often adopted to fit the
spectra of BHBs in the HSS when a power law component is rather weak. We
find that this model well reproduce the {\it Suzaku} spectra in the VHS with
$\chi^2/{\rm dof}=1266/1030$, even though the power law component
dominates the total flux at all energies from 1.2 keV to 200 keV, unlike
the case of the HSS. We obtain a power law photon index of
$\Gamma=2.79_{-0.02}^{+0.01}$ and innermost disk temperature and radius of 
$T_{\rm in} = 1.41_{-0.01}^{+0.02}$ keV and $r_{\rm in} = (24.8\pm0.5) 
\zeta_{70} d_{10}$ km from the MCD component, where $\zeta_{70} = 
(\rm{cos} \ i/\rm{cos} \ 70^{\circ})^{-1/2}$ and $d_{10} = D/(10~\rm{kpc})$. 
The model description and best-fit parameters are given in 
Tables~\ref{Model} and \ref{result}, respectively. 
Figure~\ref{4U_spec}(a) shows the residuals between the
data and model in units of $\chi$. The best-fit unabsorbed spectrum of
Model~A is shown in Figure~\ref{bestfitmodel}(a), where different
components are separately plotted. The absorbed 1.2--200 keV flux is
estimated $1.7 \times 10^{-8} \ \rm{ergs \ cm ^{-2} \ s^{-1}}$, which
corresponds to the absorbed luminosity of $2.0 \times 10^{38} \ d_{1
0} \ \rm{ergs \ s^{-1}}$ assuming isotropic emission. This high
luminosity and the dominant steep power law component confirm that the
object was in the VHS.

\begin{deluxetable*}{ll}
\tablecaption{Model description in {\footnotesize \tt XSPEC} \label{Model}}
\tablewidth{0pt}
\tablehead{
\colhead{Model} & \colhead{Descriptions in {\footnotesize \tt XSPEC}}
}
\startdata
Model A\tablenotemark{a} & {~~\tt wabs*Dscat*(diskbb+powerlaw)} \\[2pt]
Model B\tablenotemark{a,b} & {~~\tt wabs*Dscat*(simpl(D)*diskbb+simpl(C)*diskbb+}\\
 & {~~\tt~~~~~~~~~~~~~~~rdblur*rfxconv*simpl(C)*diskbb)} \\[2pt]
Model C\tablenotemark{a,b} & {~~\tt wabs*Dscat*(simpl(D)*diskbb+simpl(C)*diskbb+nthcomp+}\\
 & {~~\tt~~~~~~~~~~~~~~~rdblur*rfxconv*(simpl(C)*diskbb+nthcomp))} \\[2pt]
Model D\tablenotemark{a,b,c} & {~~\tt wabs*Dscat*(simpl(D)*dkbbfth(D)+simpl(C)*dkbbfth(D)+dkbbfth(C)+}\\
 & {~~\tt~~~~~~~~~~~~~~~rdblur*rfxconv*(simpl(C)*dkbbfth(D)+dkbbfth(C)))} \\[1pt]
\enddata
\tablecomments{Energy range is extended to 0.01--1000keV to use the convolution model}
\tablenotetext{a}{{\footnotesize \tt Dscat} is a local model, developed by \citet{Ueda10}, to take into account the effect of dust scattering}
\tablenotetext{b}{{\footnotesize \tt simpl(D)} and {\footnotesize \tt simpl(C)} represent the Direct and the Comptonized component, respectively}
\tablenotetext{c}{{\footnotesize \tt dkbbfth(D)} and {\footnotesize \tt dkbbfth(C)} represent the Disk and Compton component of the {\footnotesize \tt dkbbfth} model, respectively\\}
\end{deluxetable*}

\subsection{Disk and Corona with Reflection (Model B)}\label{modelB}

Hereafter, we apply more physically-motivated models, where the power
law component is modeled as Comptonization of seed photons emitted from
the disk. Essentially the same models are employed to fit the {\it Suzaku}
spectrum of the BHB GX 339--4 in the VHS by \citet{Tamura12}. First, we
utilize the {\tt simpl} model \citep{Steiner09} multiplied to the
{\tt diskbb} component. The {\tt simpl} model is an
empirical model in which a fraction of the photons in an input seed
spectrum is Compton-scattered into a power-law component. We find that
this model alone does not reproduce the XIS+HXD spectra, yielding
$\chi^2/{\rm dof}=1386/1030$.

To improve the fit, we further consider the reflection of the
Comptonized component from the accretion disk by using the 
{\tt rfxconv} model \citep{Kolehmainen11}. This is a convolution model that
combines the reflected spectra of \citet{Ross05} from a constant
density ionized disk with the {\tt ireflect} model \citep[a convolution
version of the {\tt pexriv} model;][]{Magdziarz95}. To
consider general relativistic smearing effect, we utilize the model
{\tt rdblur} \citep{Fabian89}. Following \citet{Tamura12}, we fix the
iron abundance to be solar, the inner and outer radius $R_{\rm in} = 6 R_{\rm g}$,
$R_{\rm out} = 10^5 R_{\rm g}$, and the power law index of emissivity
$\beta = -3$, which are very difficult to determine from the fit.
We confirm that other choices of the parameters between
$R_{\rm in} = 6 R_{\rm g} - 10^2 R_{\rm g}$ or $R_{\rm out} = 10^3
R_{\rm g} - 10^5 R_{\rm g}$ give almost the same results
within the errors.

We refer this model including the disk reflection component as
``Model~B'', whose description in the XSPEC terminology is given in
Table~\ref{Model}. We find that Model~B gives a much better fit with
$\chi^2/{\rm dof} = 1231/1028$ than that without a reflection
component. The best-fit parameters are summarized in Table~\ref{result}.
The fitting residuals and unabsorbed best-fit model are plotted in
Figures~\ref{4U_spec}(b) and \ref{bestfitmodel}(b), respectively.
\begin{deluxetable*}{llcccc}
\tablecaption{Fitting results \label{result}}
\tablewidth{0pt}
\tablehead{
\colhead{Component} & \colhead{Parameter} & \colhead{Model A} &
\colhead{Model B} & \colhead{Model C} & \colhead{Model D}
}
\startdata
{\small \tt wabs} & 
$N_{\rm H}(10^{22}{\rm cm}^{-2})$ 
& $9.38\pm0.06$ & $8.32_{-0.06}^{+0.07}$ 
& $8.46\pm0.07$ & $8.39\pm0.07$ \\[2pt] \tableline \\[-6pt]
{\small \tt diskbb} & 
$kT_{\rm in}({\rm keV})$ 
& $1.41_{-0.01}^{+0.02}$ & $1.16_{-0.01}^{+0.02}$ 
& $1.12\pm0.02$ & --- \\[3pt] 
& $r_{\rm in}(\zeta_{70}d_{10} \ {\rm km})$\tablenotemark{a}  
& $24.8\pm0.5$ & $48.2_{-1.1}^{+1.2}$ 
& $42.2_{-1.4}^{+1.2}$ & --- \\[2pt] \tableline \\[-6pt] 
{\small \tt nthcomp} & 
$kT_{\rm in}^{\rm int}({\rm keV})$ 
& --- & --- 
& --- & $1.30_{-0.02}^{+0.04}$ \\[3pt] 
~~~or  & 
$\Gamma_{\rm th}$ 
& --- & --- 
& $3.01_{-0.04}^{+0.05}$ & $2.89_{-0.04}^{+0.05}$ \\[3pt] 
 {\small \tt dkbbfth}  & 
$kT_{\rm e}({\rm keV})$ 
& --- & --- 
& $300_{-227}$ & $53_{-13}^{+10}$ \\[3pt] 
& Norm 
& --- & --- 
& $1.6_{-0.3}^{+0.2}$ & $0.073_{-0.014}^{+0.008}$ \\[3pt] 
& $r_{\rm tran}(r_{\rm in})$ 
& --- & --- 
& --- & $5.5_{-0.9}^{+0.8}$ \\[2pt] \cline{2-6}  \\[-6pt] 
& Derived $\tau$\tablenotemark{b} 
& --- & --- 
& $0.08_{-0.01}^{+0.22}$ & $0.41_{-0.07}^{+0.12}$\\[2pt] \tableline \\[-6pt]
{\small \tt PL} & 
$\Gamma_{\rm PL}$ 
& $2.79_{-0.02}^{+0.01}$ & $2.83_{-0.03}^{+0.02}$ 
& (2.1)\tablenotemark{c} & (2.1)\tablenotemark{c} \\[3pt] 
\ or & norm & 
$21.2 \pm 0.9$ & --- 
& --- & --- \\[3pt] 
{\small \tt simpl} & 
$f_{\rm PL}$ 
& --- & $0.29_{-0.05}^{+0.02}$ 
& $0.037_{-0.07}^{+0.09}$ & $0.024_{-0.015}^{+0.009}$ \\[2pt] \tableline \\[-6pt]
{\small \tt rfxconv}\tablenotemark{d} & 
$\Omega/2\pi$ 
& --- & $0.66_{-0.07}^{+0.25}$ 
& $0.87_{-0.19}^{+0.59}$ & $1.16_{-0.33}^{+1.62}$ \\[3pt] 
& $\rm{log} (\xi)$ 
& --- & $3.30_{-0.07}^{+0.06}$ 
& $3.70_{-0.22}^{+0.15}$ & $3.30_{-0.12}^{+0.05}$ \\[2pt] \tableline \\[-6pt]
& $\chi^2/{\rm dof}$ 
& $1266/1030$ & $1231/1028$ 
& $1202/1026$ & $1189/1026$ \\[2pt] \tableline \\[-6pt] 
Photon flux\tablenotemark{e} & 
$F_{{\rm disk} + {\rm th}}^{\rm photon}$ 
& --- & --- 
& $21.1$ & $21.0$ \\[3pt]
inner radius\tablenotemark{f} & 
$r_{\rm in}^{*}(\zeta_{70}d_{10}~{\rm km})$ 
& --- & --- 
& $51.5_{-1.3}^{+1.6}$ & $41.0_{-1.7}^{+0.7}$
\enddata
\tablenotetext{a}{Calculate using only {\tt diskbb} normalization}
\tablenotetext{b}{Calculated using Equation (1) in \citet{Tamura12}}
\tablenotetext{c}{Fixed in the spectral fitting}
\tablenotetext{d}{The solar abundances are assumed for all heavy elements. The inclination angle is fixed at $70^{\circ}$. The reflected component is relativistically blurred by {\tt rdblur} with fixed $R_{\rm in}^{\rm rdblur} = 6 R_{\rm G}$, $R_{\rm out}^{\rm rdblur} = 10^5 R_{\rm G}$, and $\beta = -3$ \\[-7pt]}
\tablenotetext{e}{Unabsorbed photon flux of the sum of the disk and thermal Compton component in the range of 0.01--100 keV after excluding the reflected component}
\tablenotetext{f}{The apparent innermost disk radius is calculated by using Equation (A.1) in \citet{Kubota04a} for the slab geometry\\}
\tablecomments{Errors represent 90\% confidence limits (statistical errors only). Energy range is extended to 0.01--1000keV.}
\end{deluxetable*}

\subsection{Two Comptonization Components; Thermal and Non-Thermal (Model C)}\label{modelC}

The {\tt simpl} model used in Model~B assumes no high energy cutoff, and
hence should be regarded to represent Comptonized emission by
non-thermal electrons. However, previous studies of BHBs in the VHS
suggest that the Comptonizing corona contains both thermal and
non-thermal electrons
\citep[e.g.,][]{Gierlinski99,Kubota01,Gierlinski03}. It
is not clear whether it consists of two separate components or is a
hybrid plasma whose energy distribution is Maxwellian at low energies
but a power-law at high energies. For simplicity, here we consider
the former case and leave the detailed modeling of Comptonization by a
hybrid plasma for future work. 
We note, however, that while a hybrid electron distribution can fit 
the spectrum, such a homogeneous source model cannot explain the result by 
\citet{Axelsson14}, who show that the spectrum of the fastest variability 
is different to that of the time averaged spectrum.
Accordingly, we add a thermal Comptonization model \citep{Zdziarski96,Zycki99} 
(Model~C in Table~\ref{Model}) to
Model B. As in Model~B, we include the disk reflection component of the
Comptonized emission (both thermal and non-thermal ones).

The {\tt nthcomp} model has four parameters, the seed photon
temperature, which we tie at the disk temperature of {\tt
diskbb}, the electron temperature, 
the photon index $\Gamma_{\rm th}$, and its normalization. 
We set the upper limit of the electron temperature 
to be 300 keV, as the {\tt nthcomp} model does not include full relativistic corrections.
Since we cannot constrain both $\Gamma_{\rm th}$ in {\tt nthcomp} and
$\Gamma_{\rm pl}$ in {\tt simpl}, we fix $\Gamma_{\rm pl}=2.1$, a
typical value for a non-thermal power law component seen in the HSS
\citep{Gierlinski99}. We note that the choice of $\Gamma_{\rm pl}$ is
not sensitive to the overall fitting result as our data is available up
to only 200 keV.

In the {\tt nthcomp} model, the relation among the optical 
depth of corona, the electron temperature, and the photon index 
is given by Equation (1) of \citet{Tamura12};
\begin{equation}
\tau = \frac{1}{2} \cdot \left[ \sqrt{\frac{9}{4} + \frac{3}{\Theta_{\rm e} \cdot \left( \left( \Gamma_{\rm th} + \frac{1}{2} \right)^2 - \frac{9}{4} \right) }} - \frac{3}{2} \right]
\end{equation}
where $\Theta_{\rm e} = kT_{\rm e} / m_{\rm e}c^2$.
It is derived from Equation (A1) of \citet{Zdziarski96} by assuming 
a slab geometry.
We cannot directly estimate the innermost disk radius using
the {\tt diskbb} normalization alone, since the {\tt nthcomp}
model does not conserve the photon number unlike {\tt
simpl}. We therefore use Equation (A1) of \citet{Kubota04a}, 
\begin{eqnarray}
F_{\rm dkbbfth}^{\rm photon} &= 0.0165\left[ \frac{r_{\rm in}^2{\rm cos}i}{\left( D/10 \ {\rm kpc} \right)^2} \right ] \left( \frac{T_{\rm in}}{1 \ {\rm keV}} \right)^3 \nonumber\\ 
&~~~~~~~~~~~~~{\rm photons \ s^{-1} \ cm^{-2}},&
\end{eqnarray}
to derive the true innermost radius by considering the photon-number
conservation.
Note that we replace ``$F_{\rm disk}^{\rm p} + 
F_{\rm thc}^{\rm p}2{\rm cos}i$" in Equation (A1) of \citet{Kubota04a} with 
$F_{\rm dkbbfth}^{\rm photon}$ to assume the slab corona geometry.

\begin{figure*}[t]
\begin{center}
\plottwo{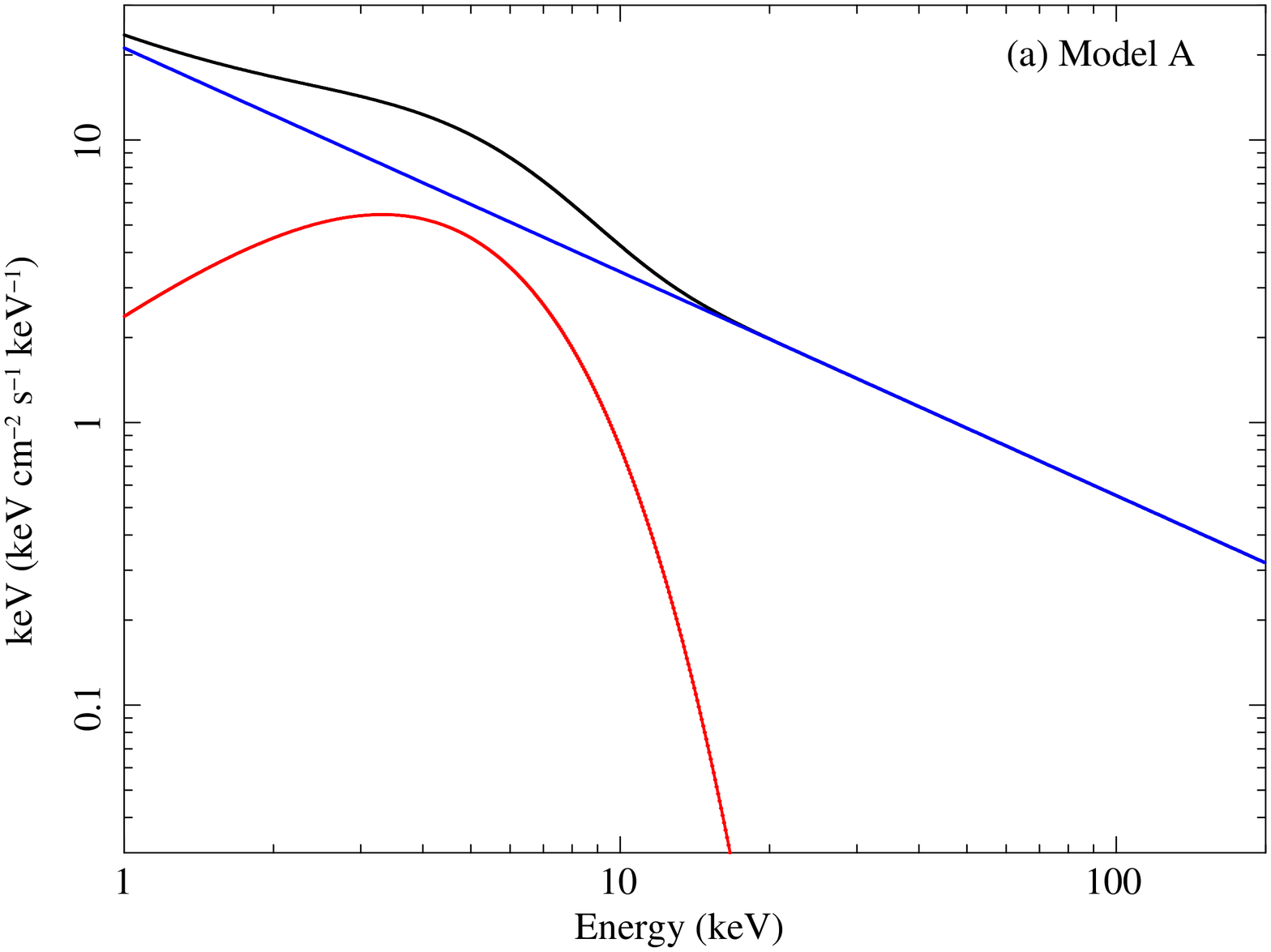}{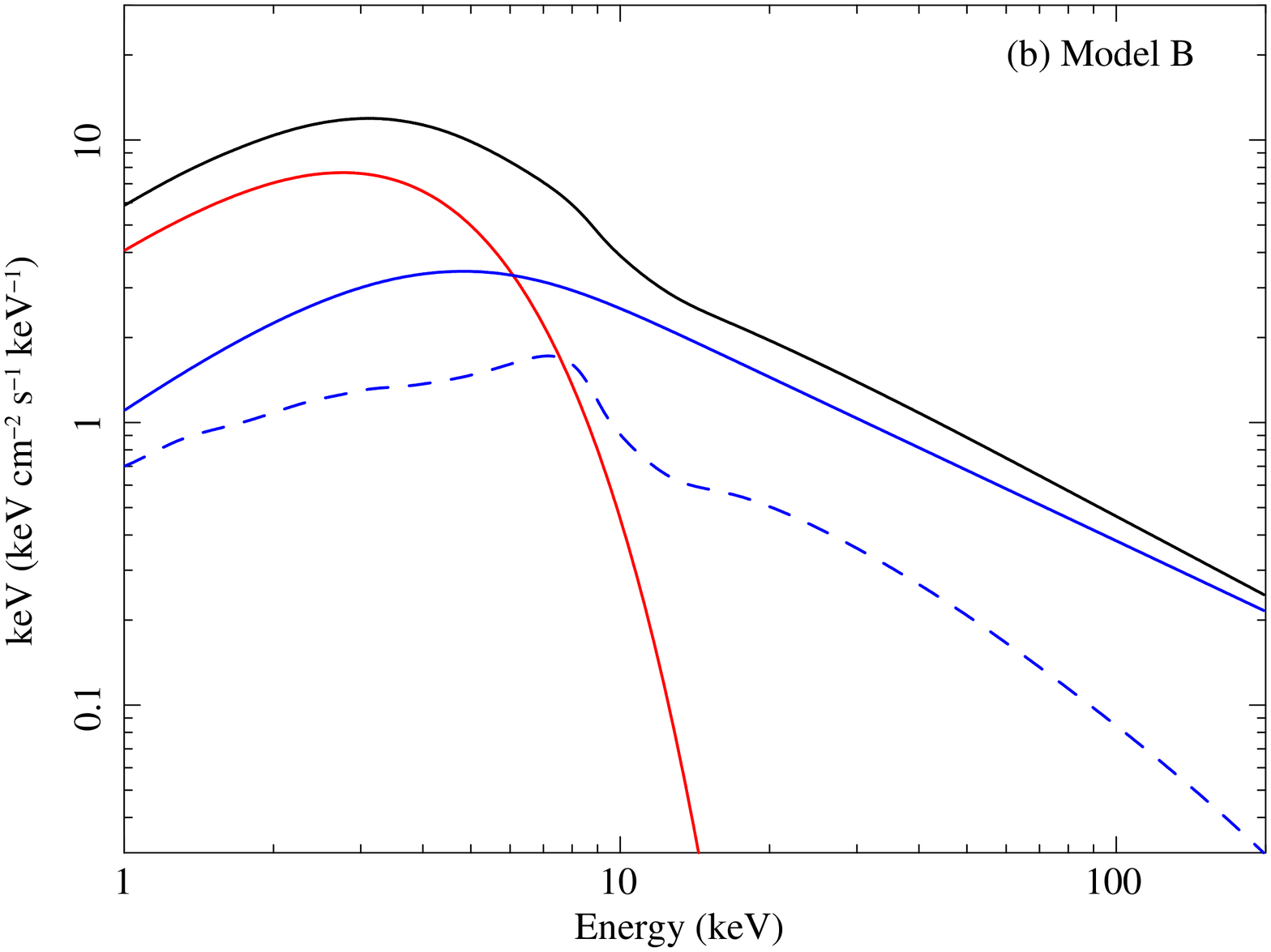}\\[8pt]
\plottwo{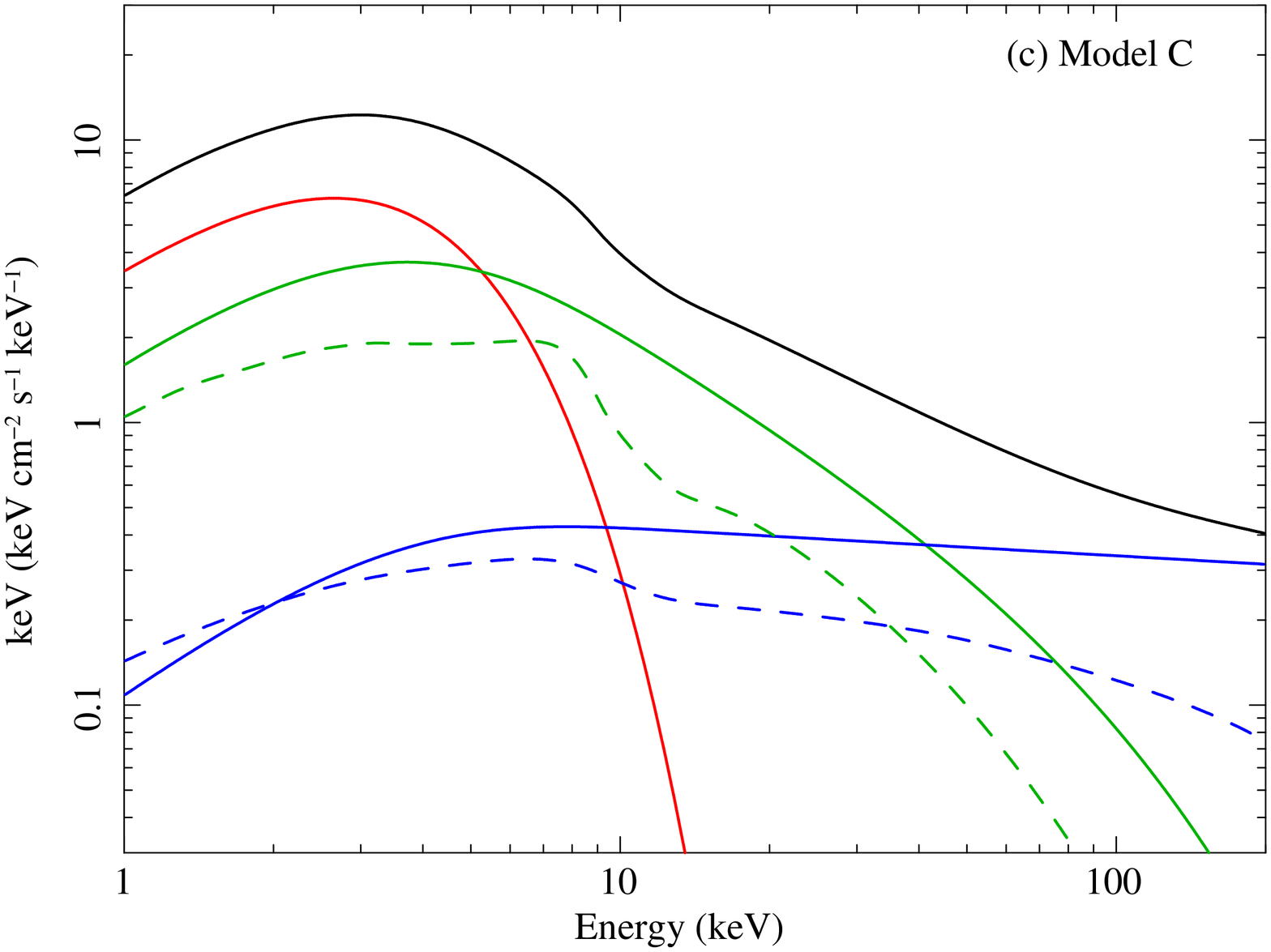}{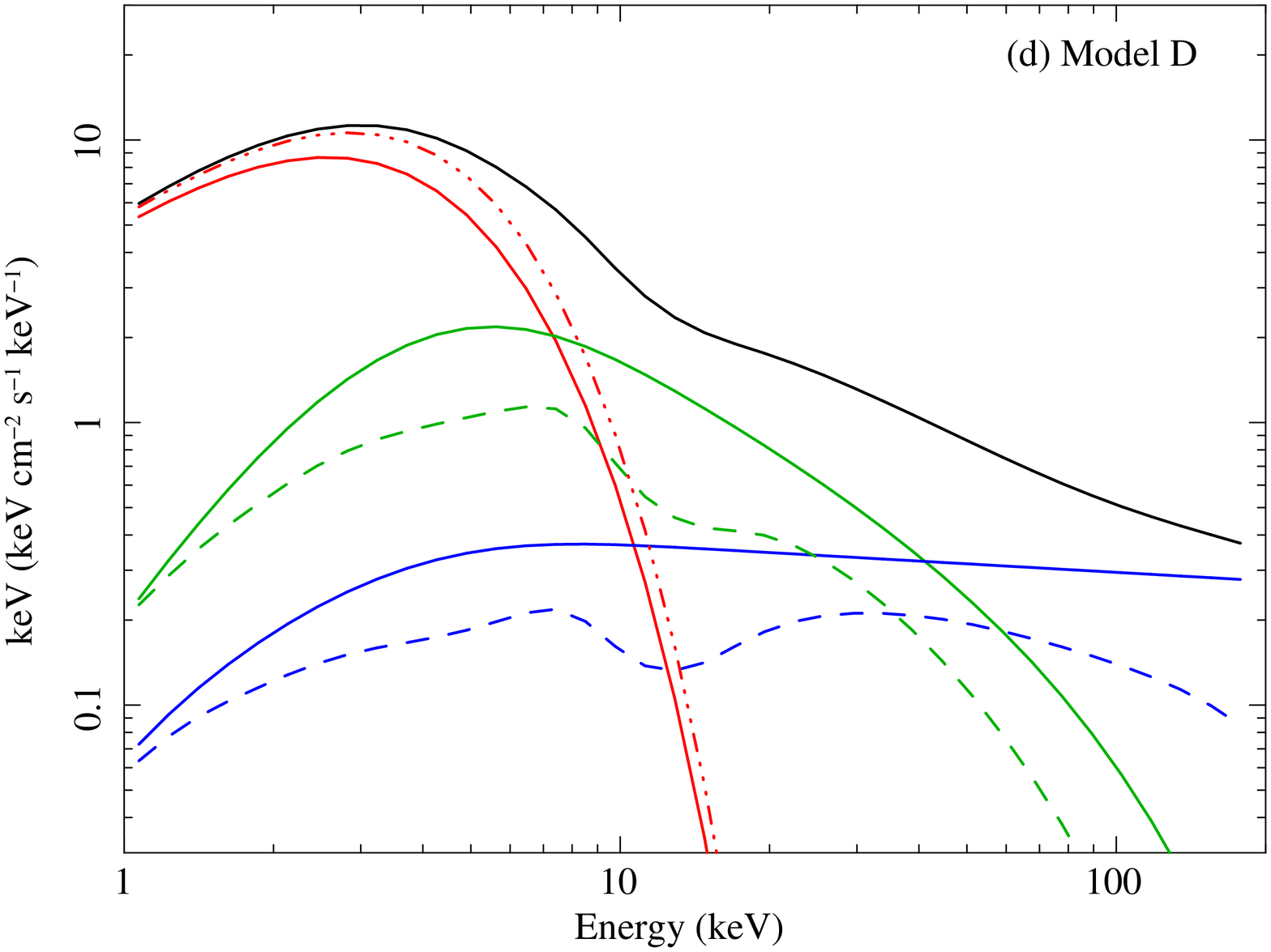}
\caption{
Best-fit model spectra corrected for the interstellar absorption
in units of $E F(E)$, where $E$ is the energy and
$F(E)$ is the energy flux, for Model~A (a), Model~B (b), Model~C (c),
and Model D (d). The observed disk component ({\tt diskbb} or {\tt dkbbfth}; red), 
non-thermal Compton component 
({\tt powerlaw} or {\tt simpl(C)}; blue), and thermal Compton component 
({\tt nthcomp} or {\tt dkbbfth(C)}; green) are separately 
plotted with the solid curves. 
The reflected non-thermal (blue) and thermal (green) Compton component
are shown with the dashed curves. In panel (d), we also plot the intrinsic
continuum (red dash-dotted line) that would be observed when all the
luminosity of the {\tt dkbbfth} component were emitted as the 
standard disk.\label{bestfitmodel}}
\end{center}
\end{figure*}

We find that Model~C gives a better fit than Model~B, yielding
$\chi^2/{\rm dof} = 1202/1026$. 
The improvement from Model B is significant at 99.9995\% 
level from an F-test, i.e. the data require
the presence of another Comptonized component at high confidence, although the
electron temperature of the {\tt nthcomp} component is not well constrained, 
with a lower limit of 73 keV.
The best fit parameters are summarized in
Table~\ref{result}. 
Figure~\ref{4U_spec}(c) plots the residuals, while
Figure~\ref{bestfitmodel}(c) shows the unabsorbed best-fit model
spectrum. It is seen that the thermal Compton component dominates below
$\sim$50 keV.

\subsection{Inner Disk-Corona Coupling (Model D)}\label{modelD}

In reality, it is expected that the hot Comptonizing corona
surrounds only an inner part of the disk.
In this picture, Model~C is unphysical because it does not
take into account any energetic coupling between the disk and
corona; the summed emissivity of the
disk blackbody and thermal Comptonization components in an inner region
is larger than the extrapolation from the outer disk that is not Comptonized
\cite[see][]{Done06,Tamura12}. This 
would lead an incorrect estimate of the innermost disk radius. To overcome
this situation, we replace {\tt diskbb} and {\tt nthcomp} to {\tt dkbbfth}
\citep{Done06} (Model D in Table \ref{Model}). The {\tt dkbbfth} model
assumes that the energy released by gravity is dissipated locally, which
is emitted purely from the standard disk at $r > r_{\rm tran}$ but is
split into the disk and corona at $r_{\rm in} < r < r_{\rm tran}$.
Thus, the inner part of the disk covered by the corona is less luminous
than it would have been if the corona did not exist. The {\tt dkbbfth}
model has five parameters, the disk temperature, the outer radius of the
corona $r_{\rm tran}$, the photon index of the Comptonized component,
the temperature of the corona, and the normalization. We set all these
parameters to be free in the fitting. Since this model only takes into
account thermal Comptonization, we further add the {\tt simpl} model to
describe non-thermal Comptonization. There we set the disk emission to
be the seed photons of {\tt simpl} and fix the photon index at 2.1.
We note that the number of free parameters in Model~D is
the same as in Model~C.

The fit shows a significant improvement ($\chi^2/{\rm dof}=1189/1026$) from Model C.
Thus, we conclude that Model D gives the best description of the
{\it Suzaku} spectra among the three physically-motivated models examined
above. The best-fit parameters are summarized in Table \ref{result} and
the fitting residuals are plotted Figure~\ref{4U_spec}(d). The
best-fit unabsorbed model is shown in Figure~\ref{bestfitmodel}(d).
The dashed-dotted curve shows the intrinsic disk
emission that would be observed when all the luminosity of the {\tt dkbbfth}
component were emitted as the standard disk.

\section{Reanalysis of XIS Spectra in the HSS}\label{analysis2}

To compare with the results in the VHS, we also perform spectral fit
to the {\it Suzaku} spectra of 4U 1630--47 taken in 2006 February 
8 when
the object was in the HSS (see Section~\ref{obs}). 
We simultaneously fit the spectra of the 4 XISs and HXD/PIN.
We utilize the energy band of 13--30 keV of the PIN spectrum.
Only the XIS spectra below 6 keV are utilized
to avoid pile-up effects that produce an artificial hard
excess above $\sim$7 keV and to ignore the strong absorption lines
from highly ionized iron ions \citep[see ][ for details]{Kubota07}.
The energy ranges of 1.7--1.9 keV and 2.2--2.4 keV are excluded, where 
calibration uncertainties are 
large. Systematic errors of 1\% and 3\% are included in each spectral
bin above 1.9 keV and below 1.7 keV, respectively. 
In the same way as for the VHS spectra, we multiply
the {\tt Dscat*wabs} model to the intrinsic continuum to take into
account dust scattering effects, which could affect the continuum
spectral parameters reported by \citet{Kubota07}. Using the {\tt
xissimargen} FTOOL, we estimate the scattering fraction
to be 0.78. The cross normalization factor of HXD/PIN relative to XIS0 is fixed
at 1.164, while those of XIS1, XIS2, and XIS3 are set to be free.
Thus, we can directly compare the absolute flux
between the two states by using the XIS0 as the reference.

We apply a model consisting of the MCD component and its
Comptonization ({\tt simpl*diskbb} in the XSPEC 
terminology).
The photon index in {\tt simpl} is fixed at 2.1
\citep[see Fig 2b of][]{Kubota07}.
Although the model is rather simple, 
the fit is found to be reasonably good with $\chi^2 / {\rm dof} =
1740/1438$. We obtain the innermost disk temperature and radius of $T_{\rm
in} = 1.31\pm0.01$ keV and $r_{\rm in} = (35.0\pm0.3) \zeta_{70} d_{10}$ km,
respectively. The derived innermost radius is larger than that reported
by \citet{Kubota07}, $r_{\rm in} \simeq 25 \zeta_{70} d_{10}$ km. 
The difference is attributable to 
the dust scattering effects. The hydrogen column density is determined
to be $N_{\rm H} = (8.30_{-0.02}^{+0.03}) \times 10^{22} \ {\rm cm^{-2}}$, which is
consistent with the \citet{Kubota07} result and our result in the VHS. 
The small scattered fraction $f_{\rm PL} = 
0.0048\pm0.0002$ means that the MCD component was dominant in this 
observation. 
The absorbed 1.2--30 keV flux is
estimated to be $6.9 \times 10^{-9}~\rm{erg \ s^{-1} \ cm^{-2}}$, which
gives an absorbed luminosity of $8.3 \times 10^{37}~{\rm erg \ s^{-1}}$
by assuming isotropic emission.

\section{Near-Infrared Observation and Results in the VHS}\label{infrared}

\subsection{Observation and Data Reduction}

We performed near-infrared photometric observations of 4U 1630--47 in the
$J$ (1 .25~${\rm \mu m}$), $H$ (1.63~${\rm \mu m}$), and $K_{\rm s}$
(2.14~${\rm \mu m}$)-bands on 2012 October 1 (one day before the {\it Suzaku}
observation) by using SIRIUS camera \citep{Nagayama03} on the 1.4-m IRSF
telescope at the South African Astronomical Observatory (SAAO).
We obtained 25 object-frames with an exposure of 15 
sec per frame, and thus the net exposure time was 375 sec.
The seeing in full width half maximum in the $J$-band was $\sim 1''.5$ 
(3.5 pixels).

We perform standard data reduction (i.e., dark subtraction,
flat-fielding, sky subtraction, and combining dithered images) with the
IRSF pipeline software on IRAF (the Image Reduction and Analysis
Facility, distributed by the National Optical Astronomy Observatory)
version 2.16. We combine all the object frames obtained in one night to
maximize signal-to-noise ratio.

The near infrared counterpart has been identified by
\citet{Augusteijn01}, but
we are not able to measure the flux by
aperture photometry since 4U 1630--47 is contaminated by nearby
stars in the infrared images. 
Thus, we perform PSF (point spread function) photometry 
using the DAOPHOT \citep{Stetson87} package in IRAF.

\subsection{Results}

\begin{figure}[t]
\begin{center}
\epsscale{1.0}
\plotone{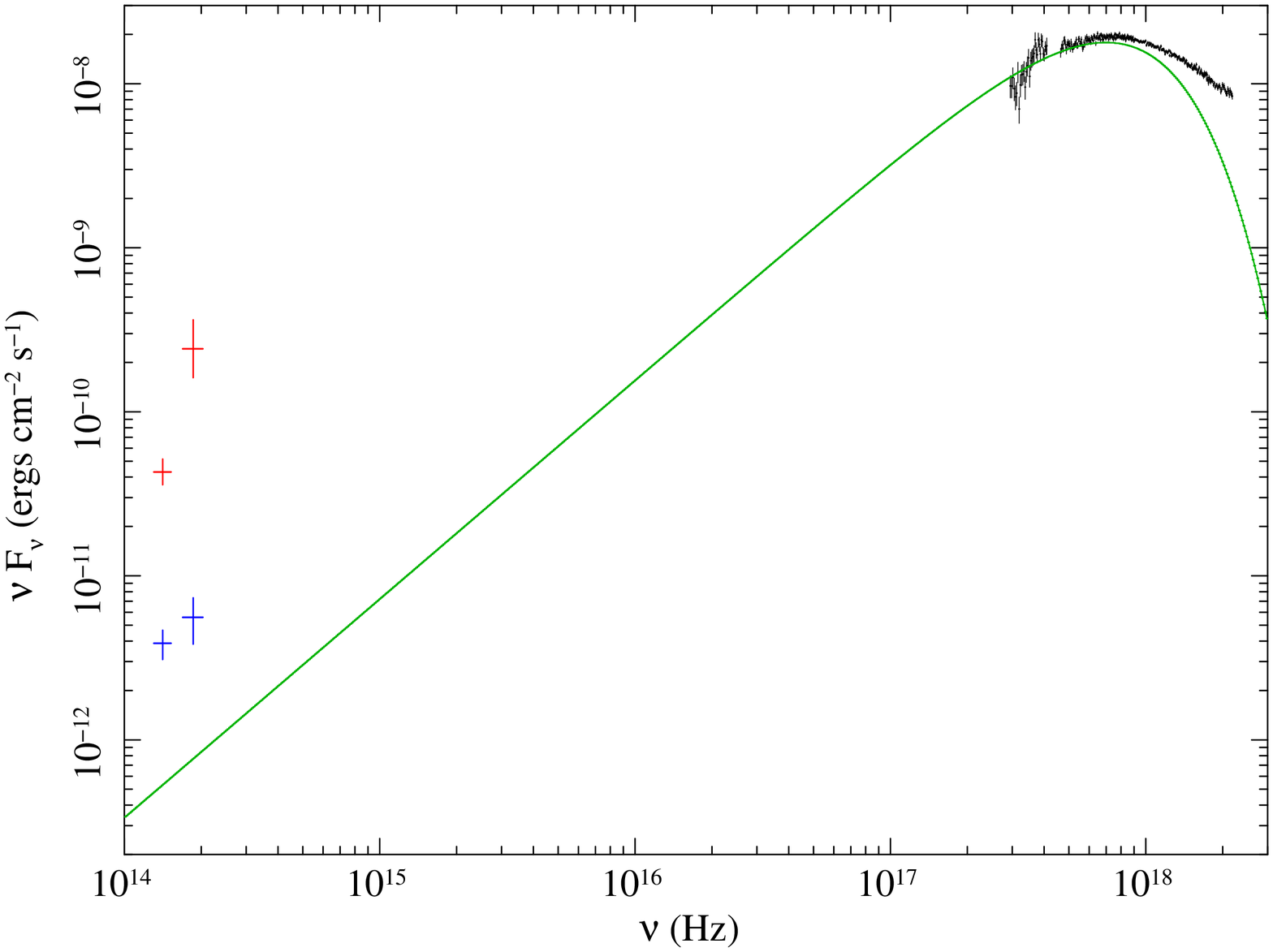}
\caption{
Spectral energy distribution of 4U 1630--47 in the VHS. The red
points correspond to the absorption corrected fluxes
in the $H(1.63~\rm{\mu m})$ and $K_{\rm s}(2.14~\rm{\mu m})$-bands obtained 
on 2012 October 1. 
These fluxes have an extremely blue color, which may 
indicate that the absorption correction is overestimated.
More moderate absorption corrected fluxes using the half of $N_{\rm H}/A_{V}$
conversion factor are also plotted in the blue color.
The exact fluxes must be located between the red points and the blue points.
The black data are the XIS0 spectrum. 
The green curve represents the intrinsic {\tt diskbb} emission that
would be observed without Comptonization in Model D.
Note that this does not match the total observed
fluxes because of the Comptonization.
All data are corrected for the interstellar extinction or absorption. 
\label{IRSFobs}}
\end{center}
\end{figure}

We obtain the near-infrared magnitudes of $17.9 \pm 0.4$ mag in the 
$H$-band and $16.0 \pm 0.2$ mag in the $K_{\rm s}$-band, while we are unable
to constrain the $J$-band flux because of the severe source confusion.
These magnitudes are almost the same as those reported in the 1998 outburst
\citep{Augusteijn01}. 
We estimate the extinction in each band as $A_H = 7.8
\pm 1.3$~mag and $A_K = 5.0 \pm 0.9$~mag by using the relation
between the hydrogen column density and optical extinction $A_V$,
$N_{\rm H} = (1.79 \pm 0.03) \times A_V [\rm{mag}] \times 10^{21}
[\rm{cm^{-2}}]$ \citep{Predehl95}, and the conversion factor from $A_V$
into $A_H$ or $A_K$ given by \citet{Rieke85}. 
The extinction-corrected IRSF fluxes ($m_H^{\rm cor} = 10.1$~mag, $m_K^{\rm cor} = 11.0$~mag) are plotted as the red points in 
Figure~\ref{IRSFobs}. Note that the 
extremely blue color ($m_H^{\rm cor} - 
m_K^{\rm cor} = -0.9$~mag),
which is hard to explain by any stellar or continuum source,
most probably 
comes from the errors in the photometry, $N_{\rm H}$, and/or
the $N_{\rm H}$ to $A_V$ conversion.
We find that the last one can produce the largest uncertainty.
Figure~3 of \citet{Predehl95} shows that there is a deviation 
by a factor of $\sim$2 in 
the relation between $N_{\rm H}$ and $A_V$ for some objects. 
Accordingly, we consider an extreme case
where the $A_V$ to $N_{\rm H}$ ratio is twice smaller than the above as
a lower limit of the extinction correction.
This gives the corrected 
fluxes of $m_H^{\rm cor} = 13.8$~mag and $m_K^{\rm cor} = 13.4$~mag, 
which are plotted as the blue points in Figure~\ref{IRSFobs}. 
Thus, the extinction-corrected magnitudes are estimated to be in the
range of $13.8~\rm{mag} > m_H^{\rm cor} > 10.1~\rm{mag}$ and 
$13.4~\rm{mag} > m_K^{\rm cor} > 11.0~\rm{mag}$.

In Figure \ref{IRSFobs}, the intrinsic thermal emission from the
standard disk based on Model D (Section~\ref{modelD}) that would be
observed without Comptonization (i.e., the red dashed line in
Figure~6(d)) is plotted by the green solid curve.
As noticed, even the lower limits of
the observed near-infrared fluxes are larger than the
expected fluxes from the MCD component.
This suggests that the fluxes are
dominated by other components. 
We estimate the absolute magnitudes to be
$-1.2~\rm{mag} > M_H^{\rm cor} > -5.3$ mag and 
$-1.6~\rm{mag} > M_K^{\rm cor} > -4.2$ mag for a 
distance of 10 kpc. Thus if the near-infrared fluxes completely come 
from the companion star, it would be an early type star (O7-B2), supporting 
the suggestion by \citet{Augusteijn01}, or a K type giant star.

These magnitudes should be regarded as the upper limit of the true
luminosity of the companion star, however, as we also expect other
origins that could be responsible for the near-infrared fluxes, 
including
(1) the irradiation of the outer disk by the X-ray emission,
(2) synchrotron radiation from the jets, and/or
(3) cyclo-synchrotron emission from the hot flow \citep{Veledina13}.
Unfortunately, the large uncertainties in the extinction-corrected
fluxes make it very difficult to discriminate these possibilities.
We note that the observed (uncorrected) near-infrared fluxes in the VHS
are almost the same as those reported by \citet{Augusteijn01} when the
source was in the HSS with a similar X-ray flux level. Since jet
ejection is believed to be quenched in the HSS, this comparison suggests
that the synchrotron emission from the jets, if any, is not a dominant
source of the near infrared fluxes in the VHS. 

\section{Discussions}\label{discuss}

\subsection{Result Summary}

Using {\it Suzaku}, we have obtained so-far the best quality broad
band X-ray spectra of 4U 1630--47 in the VHS, covering the 1--200 keV
band with good energy resolution by X-ray CCD below 10 keV. The
observation epoch corresponds to when the hard X-ray flux above 15 keV
was peaked during the 2012 September-October outburst. In the spectral
analysis, we carefully take into account the effects by dust
scattering.

The continuum is found to be well approximated by a phenomenological
model consisting of a MCD and a power law component dominating the
total flux in the entire 1.2--200 keV band (Model A). However, we
regard that this modeling is unnatural because (1) no low energy
turn-over is included, which should exist if the power law component
is originated by Comptonization of seed photons emitted from the disk,
and (2) the derived column density is significantly larger than that
obtained in the HSS. We thus consider three physically-motivated
models where the disk photons are Compton-scattered by a hot corona
surrounding it. Among them, Model D, where the Comptonizing corona
is energetically coupled with the underlying standard disk below a
transition radius, best reproduce the spectra. A significant
reflection component from the inner disk with a solid angle of
$\Omega/2\pi \approx 1.2$ is detected. 
This is consistent with the recent result by \citet{King14}, although 
the state was different from ours.
Below, we discuss the inner
disk geometry on the basis of the result of Model D. No significant
emission or absorption lines are detected in our spectra. This fact
gives tight constraints on the presence of the baryonic jets
discovered by \citet{Diaz Trigo13} 4 days before the {\it Suzaku}
observation, and that of the disk wind seen in the HSS
\citep{Kubota07,King14}.

\subsection{Inner Disk Truncation in the VHS}\label{truncation}

We find that the standard disk in the VHS is most likely truncated 
before reaching the ISCO. From Model~D, the best physically
self-consistent model considered here, we derive the innermost radius of
the standard disk is $r_{\rm in} = 41.0_{-1.7}^{+0.7} \zeta_{70}
d_{10}$ km in the VHS. This is significantly larger than that found in
the HSS, $r_{\rm in} = (35.0\pm0.3) \zeta_{70} d_{10}$ km, which is
found to be constant during this state and hence most probably
corresponds to the ISCO \citep{Kubota07}. Since we directly compare the
results obtained with the same detector (XIS0), the comparison is free
from any cross-normalization uncertainties between different
instruments. Also, a common model is employed to represent the
interstellar absorption and dust scattering, making the relative
systematic error least. 
When we assume an extreme case that the soft X-ray flux 
producing the dust scattering halo is by 10\% lower
than the Suzaku epoch (see Section~\ref{dust}), we obtain an even larger 
innermost radius. Thus, the conclusion of disk truncation is robust against 
uncertainty in the correction for dust scattering.

A truncated disk in the VHS is also suggested from the BHB GX 339--4
by applying essentially the same model as our Model~D
\citep{Tamura12}. Hence, disk truncation would be a common feature of
the VHS, although the innermost radius is less than twice that of the
ISCO in both objects. We note that the ``true'' disk radius, $R_{\rm
in}$, is estimated as $R_{\rm in} = \kappa \xi^2 r_{\rm in}$, where
$\kappa$ is the color-to-effective temperature ratio and $\xi$ is a
correction factor for the stress-free boundary condition at the ISCO
\citep{Kubota98}. To explain the $r_{\rm in}$ value in the VHS by the
ISCO, $\kappa$ must be smaller than that in the HSS. As discussed in
\citet{Tamura12}, this case is quite unlikely under strong irradiation
by the corona in the VHS.

\subsection{Life Time of Baryonic Jets}\label{jet}

Relativistic baryonic jets of 4U 1630--47 were detected with {\it
XMM-Newton} on 2012 September 28 \citep{Diaz Trigo13} when the source
was also in the VHS. The {\it XMM-Newton} spectrum shows
Doppler-shifted Fe XXVI lines at 7.3 keV and 4 keV. Assuming that they
are twin jets of the same physical parameters emitted to opposite
directions, they estimate the jet speed, the inclination angle, and
the cone angle to be $0.3-0.4c$, $58^{\circ}-67^{\circ}$, and
$3^{\circ}.7-4^{\circ}.5$, respectively. The parameters are quite
similar to those of SS~433 \citep{Namiki03}, the unique Galactic
microquasar exhibiting steady relativistic jets with a velocity of
0.26$c$, which most likely contains a black hole \citep{Kubota10}.

In contrast, our {\it Suzaku} spectra, taken only four days after the
{\it XMM-Newton} observation, do not show such emission-line
features. The equivalent width of the 7.3 keV emission line detected
by \citet{Diaz Trigo13} is constrained to be $<$3 eV (90\% confidence
limit), which is about 1/12 of that in the {\it XMM-Newton}
spectra. This suggests that the relativistic 
jets detected by \citet{Diaz
Trigo13} had vanished in four days. In the case of SS 433, the length
of the X-ray emitting region of the jets is $\sim 10^{13}$ cm,
corresponding to a traveling time of only $\sim 10^3$ s
\citep{Kotani96}. Thus, if we assume the similar conditions as SS 433,
the jets of 4U 1630--47 would vanish in an hour after the ejection
stops.
We cannot, however, exclude the possibilities
that a jet might still exist, such as compact jets
often seen in the LHS \citep{Fender04}.

The variation of the X-ray flux can provide clues to reveal the
condition for the ejection of baryonic jets from a black hole.
According to the {\it MAXI} and {\it Swift}/BAT light curves 
(Figure~\ref{MAXI_BAT}), 
the {\it Suzaku} observation was performed just after the hard X-ray
flux showed a sudden increase from MJD$\simeq$56201. The same trend is
confirmed in the soft band, though less clear because of the data gap
of {\it MAXI}. The source was by $\sim$30\% fainter when the jets were
detected with {\it XMM-Newton} (MJD=56198) than in the {\it Suzaku}
epoch (MJD=56202). Thus, phenomenologically, it would be possible that
the launch of the baryonic jets somehow ceased as more accretion power
was dissipated into radiation.
However, we estimate the kinetic
power of the jets to be $> 10^{40}$ erg s$^{-1}$ on the basis of the
jet model of SS~433 \citep{Kotani96}, assuming that it is proportional to
the luminosity of the ionized iron-K emission lines. This huge power
cannot be accounted for by the observed increase in the luminosity at
MJD$\simeq$56201 ($\Delta L_{\rm X} \sim 10^{38}$ erg
s$^{-1}$) unless the radiation efficiency is very low. Thus, a
significant decrease in the intrinsic mass accretion rate would be
required. To obtain firm observational clues on the jet formation
mechanism, we need more frequent monitoring observations of BHBs in high 
luminosity states with high-quality X-ray spectroscopy.

\subsection{Disk Wind in the VHS}\label{jet}

The {\it Suzaku} spectra of 4U 1630--47 in the VHS do not show any
significant absorption line features of highly ionized iron ions that
were detected in the HSS by \citet{Kubota07}. The upper limits for the
equivalent widths of the absorption lines are 1.3 eV and 1.4 eV (90\%
confidence limits) at 6.72 keV and 6.99 keV, corresponding those of
\ion{Fe}{25} and \ion{Fe}{26} observed in the HSS, respectively. There
are two possibilities to account for this fact; (1) there is actually
no disk wind in the VHS or (2) it exists but simply become invisible
by being almost fully ionized by the strong X-ray irradiation.

To examine the latter possibility, we perform simulations of a
photo-ionized gas using XSTAR version 2.2.1bk. Following \citet{Kubota07}, 
we first reproduce the physical parameters (the ionization stage and column
density) of the disk wind detected in the HSS by assuming the same
spectrum and luminosity as observed. Then, we change the X-ray
spectrum and luminosity to those in the VHS by keeping the other
parameters same. 
We find that the ion fractions of \ion{Fe}{25} and
\ion{Fe}{26} becomes $\approx$1/20 and $\approx$1/4 of those in
the HSS. The predicted equivalent width of \ion{Fe}{26} is larger than
the observed upper limit, suggesting that photo-ionization effects
alone cannot explain the absence of the absorption lines in the 
{\it Suzaku} spectra. 
However, it is still plausible that a disk wind with a smaller
density (and/or at a smaller distance) than that in the HSS is present in the VHS 
(see \citealt{Chakravorty13} for discussion of thermodynamical stability, 
and \citealt{Begelman83} for discussion of launch radius and density). Further observations
of BHBs with good energy resolution covering different states and luminosities are necessary
to establish the critical conditions to launch a disk wind.

\subsection{Conclusion}\label{conclusion}

We observed 4U 1630--47 in the VHS with {\it Suzaku} and IRSF during 
the 2012 September-October outburst.

The conclusions are summarized as follows:

\begin{enumerate}

\item The time-averaged {\it Suzaku} spectra are well described by a
physically self-consistent model consisting of thermal and non-thermal
Comptonization of the inner disk emission with energetic 
coupling between the disk and corona. A strong relativistic reflection
component from the accretion disk is required. 
We properly take into account the effects of dust scattering.

\item Comparing the result of the {\it Suzaku} spectra in the HSS observed in
2006 February, we find evidence that the accretion disk
is slightly truncated before reaching to the ISCO in the VHS.

\item Our spectra do not show any Doppler-shifted line emissions from
the relativistic jets that were detected four days before our
observation with {\it XMM-Newton}. This suggests that the jets were a
transient event and vanished between these observations.

\item 
We do not detect absorption line features from highly 
ionized iron ions that were previously present in the high/soft state. 
The upper limits on the equivalent width of these lines are not compatible 
with over-ionization by the higher luminosity irradiation of same wind. 
If a wind still remain, it
has changed in launch radius and/or density from that seen in the HSS.

\item From near-infrared data taken with IRSF one day before the {\it Suzaku}
observation, we obtain $17.9\pm0.4$ mag in the $H$-band and 
$16.0\pm 0.2$ mag in the $K_{\rm s}$-band 
for the counterpart. Since the extinction-corrected fluxes are
larger than the extrapolated fluxes from the MCD component, contribution
from a companion star, which would be 
an early type star or a K-type giant star if it dominates the
total fluxes, and/or other origins such as irradiation of the outer disk by the X-ray
emission are required.

\end{enumerate}

\acknowledgements

We are grateful to the {\it Suzaku} project for performing this ToO
observation upon our request. We thank 
Tatsuhito Yoshikawa for his help on the analysis of the
IRSF data, and Shin'ya Yamada and Shin Mineshige
for discussions. This work was partly supported by the
Grant-in-Aid for Scientific Research No.23540265 (Y.U.), 
No.24111717 (K.T.), and for young researchers (M.S.).

\end{document}